\documentclass[]{interact}

\usepackage{epstopdf}
\usepackage[caption=false]{subfig}

\usepackage[natbibapa,nodoi]{apacite}
\usepackage{chemformula} 
\usepackage[T1]{fontenc} 
\usepackage{array}
\usepackage{amsmath}
\usepackage{graphicx}
\theoremstyle{plain}

\theoremstyle{definition}

\theoremstyle{remark}

\begin{document}


\title{Resolving issues of scaling for gramian based input-output pairing methods} 

\author{
\name{Fredrik Bengtsson \textsuperscript{a}\thanks{CONTACT Fredrik Bengtsson. Email: fredben@chalmers.se}, Torsten Wik \textsuperscript{a} and Elin Svensson\textsuperscript{b}}
\affil{\textsuperscript{a}Department of Electrical Engineering, Chalmers University of Technology, SE 412 96 G{\"o}teborg, Sweden. \textsuperscript{b}CIT Industriell Energi AB, Chalmers Teknikpark, Sven Hultins gata 9D, SE 412 58 G{\"o}teborg, Sweden.}
}
\maketitle

\begin{abstract}
A key problem in process control is to decide which inputs should
control which outputs. There are multiple ways to solve this problem,
among them using gramian based measures, which include the Hankel
interaction index array, the participation matrix and the $\Sigma_{2}$
method. The gramian based measures however have issues with input and
output scaling. Generally, this is resolved by scaling all inputs and outputs
to have equal range. However, we demonstrate how this can result in
an incorrect pairing and examine alternative methods of scaling the gramian based measures,
using either row or column sums, or by utilizing the Sinkhorn-Knopp
algorithm. The benefits of these scaling strategies are first illustrated
by applying them to the control structure selection for a heat exchanger
network. Then, to more systematically analyze the benefits of the
scaling schemes, a multiple input multiple output model generator
is used to test the different schemes on a large number of systems.
This, along with implementation of automatic controller tuning, allows
for a statistical comparison of the scaling methods. This assessment
shows considerable benefits to be gained from the alternative scaling
of the gramian based measures, especially when using the Sinkhorn-Knopp
algorithm. The use of this method also has the advantage that the
results are completely independent of the original scaling of the
inputs and outputs.
\\\resizebox{25pc}{!}{\includegraphics{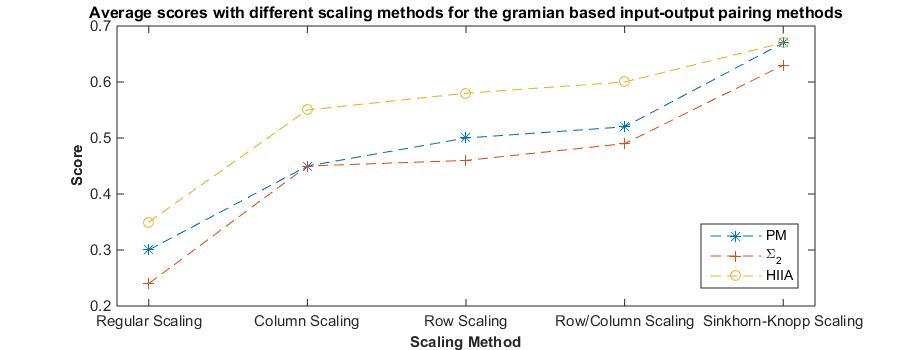}}
\\\textbf{Word Count:} 4712

\end{abstract}


\section{Introduction}

A common issue in many industrial process control systems is that interaction between different parts of the plant gives rise to a multiple input multiple output (MIMO) system, where the same input may affect multiple outputs, or conversely, the same output is affected by multiple inputs. This is the core of the input-output pairing problem; which control variables should be used to control which process parameters. While one often solves this by matching one input to one output by a decentralized configuration, at times it can be necessary to add additional feed-forward between the inputs, or even implementing MIMO controllers for parts of the system. 

There are numerous proposed input-output pairing methods, many of
which are discussed by for example, \cite{vandeWal2001487}. The most
widely used is probably still the Relative Gain Array (RGA) \citep{bristol1966new}
and modifications of it, such as the dynamic RGA and the Relative
Interaction Array (RIA) \citep{zhu1996variable}. Relatively recently
a new group of input-output pairing methods have been introduced,
namely the gramian based methods. This group includes the $\Sigma_{2}$
method \citep{birk2003note}, the participation matrix (PM) \citep{914730}
and the Hankel interaction index array (HIIA) \citep{wittenmark2002hankel}.
These methods use the controllability and observability gramians to
create an interaction matrix which gives a gauge of how much each
input affects each output. An attractive property of these interaction
matrices is that they can be used to determine both a decentralized
controller structure and a sparse structure (a structure which includes
feed-forward or MIMO blocks). Moreover, the gramian based measures
take into account system dynamics and not only the steady state properties. 

The gramian based methods, however, differ from the RGA and its variants
in that they suffer from issues of scaling, in the sense that the
results of the methods vary depending on input and output scaling.
There is a commonly suggested method to solve this problem, presented
by for example, \cite{salgado2004mimo}. We will however demonstrate on a heat exchanger network how this method can be insufficient.  \cite{arranz2009new}
have presented a different method to scale the $\Sigma_{2}$ interaction matrix 
and here we will examine this method in more detail and also apply
it to the PM and the HIIA. Furthermore, we will introduce and examine
a new method of scaling, based on the Sinkhorn-Knopp algorithm  \citep{sinkhorn1967concerning}. 

To demonstrate the benefit of the new scaling schemes a MIMO model
generator will be used. This allows us to generate a large number
of systems with predefined statistical properties, which we use for
a more general comparison of the different scaling methods.  We show that considerable improvements are made with the different scaling schemes, especially when scaling using the method of scaling based on Sinkhorn-Knopp algorithm.

The article is structured as follows: in Section 2 the different scaling methods that will be used are explained. In Section 3 the controller design strategies that will be used are presented, while in Section 4 a example system is used to demonstrate the need for new scaling methods for the gramian based measures. In Section 5 we present an evaluation of the scaling methods. Finally in Section 6 the result is summarised and possible expansions are discussed.

\section{Gramian based interaction measures, modifications and implementation }

\subsection{Gramian based measures}

The gramian based measures (PM, HIIA and $\Sigma_{2}$) can be calculated
from a system's transfer function matrix (TFM) \citep{birk2003note,914730,wittenmark2002hankel}.
Given a TFM

\[
G(s)=\begin{bmatrix}g_{11}(s) & g_{12}(s) & \cdots & g_{1n}(s)\\
g_{21}(s) & g_{22}(s)\\
\vdots &  & \ddots\\
g_{n1}(s) &  &  & g_{nn}(s)
\end{bmatrix}
\]
each measure generates an interaction matrix (IM). For the HIIA and
$\Sigma_{2}$ it is generated by

\[
[\Gamma]_{ij}=\frac{||g_{ij}(s)||}{\sum_{kl}||g_{kl}(s)||}
\]
using the Hankel norm and 2-norm for the HIIA and $\Sigma_{2}$, respectively.
The PM is derived in a similar fashion, but it uses the squared Hilbert-Schmidt
norm, i.e. the IM is generated by

\[
[\Gamma]_{ij}=\frac{||g_{ij}(s)||_{HS}^{2}}{\sum_{kl}||g_{kl}(s)||_{HS}^{2}}.
\]

Once an IM is generated, a decentralized pairing is generated by choosing
the pairing that yields the largest sum of elements from the IM. For
efficient implementation in finding which pairing yields the largest
sum of elements one can for example use the hungarian algorithm \citep{fatehi2011automatic}. 

\subsection{Scaling of the IMs}

An issue with these three methods is that the interaction matrix will
be effected by the scaling of the inputs and outputs such that different
scalings may yield different results. Generally, this is handled
by scaling the input and outputs to range 0 to 1, setting zero to the
lowest value they are likely to reach and 1 to the highest value \citep{salgado2004mimo}.
However, this scaling is at times insufficient, and we will present
a few ways in which the IMs could be rescaled for improved results. 

\subsubsection{Row or Column scaling}

Each column in the IM corresponds to the interactions from one input,
while each row corresponds to the interactions affecting one output.
If one column contains significantly less interaction than the other
columns (as may be the case if one input is relatively poorly suited
for control), little importance will be given to the decision of which
output should be controlled by this input. This may lead to a poor
input-output pairing as will be demonstrated with an example in Section
4. One way to resolve this would be to normalize the columns, that
is to divide the elements in each column of the IM by the corresponding
column sum. This ensures that, when conducting the pairing algorithm,
equal importance is given to each input. In the new IM the scaled
elements would become

\[
[\Gamma_{c}]_{ij}=\frac{[\Gamma]_{ij}}{\sum_{k=1}^{N}[\Gamma]_{kj}},
\]
where $\Gamma_{c}$ is an interaction matrix with normalized columns.
If we instead wish to ensure that equal importance is given to each
output, we can instead normalize the rows, which gives a interaction
measure defined by

\[
[\Gamma_{r}]_{ij}=\frac{[\Gamma]_{ij}}{\sum_{k=1}^{N}[\Gamma]_{ik}}.
\]

\subsubsection{Choosing between row and column scaling}

It may be difficult to determine if it is preferable to scale by rows
or columns. We propose an approach to scaling that tries to determine
which is the most appropriate for a given IM. In this approach the
column sums and row sums were first calculated. If the smallest sum
is a row sum, then the rows are scaled, and otherwise the columns
are scaled. 

\subsubsection{Sinkhorn-Knopp algorithm} \label{SKalgo}

By scaling the columns or rows we can guarantee that equal importance
is given to either each input or each output when determining pairing.
If we, however, wish to have both the columns and rows scaled we can
use the Sinkhorn-Knopp algorithm. This algorithm combines row and
column scaling by alternating between normalizing the rows and normalizing
the columns. In cases where the matrix can be made to have positive
elements on the diagonal (as is always the case with gramian based
measures) this algorithm is guaranteed to converge to a matrix that
will have both rows and columns normalized \citep{sinkhorn1967concerning}.
While the Sinkhorn-Knopp algorithm can be implemented by simply alternating
between dividing the elements in each column of the IM by the corresponding
column sum and dividing the elements in each row by the corresponding
row sum, it can also be implemented as described by \cite{knight2008sinkhorn}, i.e.

\begin{eqnarray*}
r_{0} & = & e\\
c_{k+1} & = & \mathcal{D}(\Gamma^{T}r_{k})^{-1}e\\
r_{k+1} & = & \mathcal{D}(\Gamma c_{k+1})^{-1}e,\\
\Gamma_{SK} & = & \mathcal{D}(r)\Gamma\mathcal{D}(c).\\
\epsilon_{k} & = & ||c_{k}\circ\mathcal{D}(c_{k+1})^{-1}-e||_{1},
\end{eqnarray*}
where $\circ$ denotes element-wise multiplication, $e$ is a vector
of ones, and $\mathcal{D}(x)$ turns a vector into a diagonal matrix
by creating a matrix with the elements of the vector on its diagonal.
$\epsilon_{k}$ is how far the IM ($\Gamma_{SK})$ is from being perfectly
scaled (that is having both column and row sums of one), which can
be used as a stopping criterion. 

Scaling the IMs with the Sinkhorn-Knopp algorithm has the additional
benefit of removing the impact of input and output scaling on the
IMs. Using the Sinkhorn-Knopp algorithm to scale the system will yield
the same IM, regardless of what the original scaling of the system
was.  

While the Sinkhorn-Knopp algorithm ensures that all inputs and outputs
are given equal importance, this is not necessarily what is desired.
Some outputs may be particularly important to control well. However,
as the Sinkhorn-Knopp algorithm normalizes the entire IM, it can be
used to establish a baseline to which further scaling can be done.
After scaling using the Sinkhorn-knopp algorithm the user can increase
the emphasis on finding a good match for a specific output or input,
by multiplying their respective column or row by a factor larger than
one.

\subsection{Niederlinski Index}

The Niederlinski Index (NI) can be used to determine a necessary condition
for a decentralized closed loop system to be stable \citep{doi:10.1021/i100018a015}.
Consider a system described by a TFM $G(s)$ controlled by a decentralized
and diagonal controller $C(s)$ with integral action. If $G(s)$ is
stable, $G(s)C(s)$ is proper, and all SISO control loops (created
by opening the other loops) are stable, a necessary condition for
the existence of a stable control scheme with integral action is

\[
NI=\frac{\det[G(0)]}{\prod_{i=1}^{n}g_{ii}(0)}\geq0,
\]
where $g_{ii}(0)$ refers to the diagonal elements of $G(0)$. Here,
we will use the NI in combination with the gramian based method. That
is to say that we will discard solutions which have a negative NI,
even if they have the highest total interaction, and instead choose
the solution with the highest interaction among those that have a
positive NI. 

\subsection{Sparse Controller}

The gramian based IMs can also be used to generate a sparse controller.
To do this we first start by deriving the pairing for the decentralized
controller, as described previously. Then the system is examined for
the possibility to use decoupling feed-forward. To understand how
this works, we begin by examining a 3 by 3 system, i.e.

\[
\begin{bmatrix}y_{1}\\
y_{2}\\
y_{3}
\end{bmatrix}=\begin{bmatrix}G_{11}(s) & G_{12}(s) & G_{13}(s)\\
G_{21}(s) & G_{22}(s) & G_{23}(s)\\
G_{31}(s) & G_{32}(s) & G_{33}(s)
\end{bmatrix}\begin{bmatrix}u_{1}\\
u_{2}\\
u_{3}
\end{bmatrix}.
\]

Let us assume that the inputs and outputs have been ordered such that
our decentralized controller design decided on a diagonal pairing
where $y_{i}$ is controlled by $u_{i}$ for $\forall i$. Now, $u_{1}$
will also affect $y_{2}$ and $y_{3}$ by $G_{21}(s)$ and $G_{31}(s)$,
respectively. If $u_{1}$ affects $y_{3}$ to such an extent that
it poses a problem, this can ideally be resolved by using the feed-forward

\begin{equation}
u_{3}=u_{3}^{*}-\frac{G_{31}(s)}{G_{33}(s)}u_{1},
\end{equation}
where $u_{3}^{*}$ is the control signal from the decentralized controller
and we assume $\frac{G_{31}(s)}{G_{33}(s)}$ is stable and proper.
If we implement this feed-forward loop we will have removed the direct
effect of $u_{1}$ on $y_{3}$. However, there are other consequences
of this implementation since the change of $u_{3}$ will also affect
$y_{1}$ and $y_{2}$. If these interactions are significant the feed-forward
loop might do more harm than good. Having this in mind, we examine
how the IM can be used to determine when feed-forward might be appropriate.
Consider an interaction matrix 
\[
\Gamma=\begin{bmatrix}\gamma_{11} & \cdots & \gamma_{1N}\\
\vdots & \ddots & \vdots\\
\gamma_{N1} & \cdots & \gamma_{NN}
\end{bmatrix}.
\]

First we choose the elements for the decentralized pairing as described
previously and assume, without loss of generality, that the pairing
elements are on the diagonal. After this, we look in the interaction
matrix for large elements not yet selected for pairing. However, implementing
feed-forward on the corresponding inputs needs to be weighed against
other potential interactions. For example, assume that $\gamma_{N1}$
is a large value and thus $u_{1}$ is a potential candidate for feed-forward.
However, as described in the example, this will impact $u_{N}$, which
will not only impact $y_{N}$, but also the other outputs. A gauge
of the size of this impact is $\sum_{i=1}^{N-1}\gamma_{iN}$. If these
values are very large then the IM indicates that adding the described
feed-forward on $u_{1}$ is unwise. To determine the use of feed-forward
in the general case we therefore create a new matrix $IM^{*}$, whose
elements are defined by 
\[
\gamma_{ij}^{*}=\gamma_{ij}-\rho\sum_{\substack{k=1\\
k\neq i
}
}^{N}\gamma_{ki},
\]
 where $\rho$ is a tuning parameter. With this new IM, the largest
elements where $i\neq j$ are chosen for feed-forward until the sum
of elements chosen (both for control and feedforward) is larger than
0.7, a rule of thumb for gramian based measures \citep{salgado2004mimo}.
However, as feed-forward increases controller complexity it is only
implemented if it seems likely that it will have a positive impact.
This is determined by checking if $\gamma_{ij}^{*}>0$ in which case
feed-forward is considered appropriate, and otherwise it is not implemented.
Further precautions also have to be taken to avoid implementing an
unstable or non-proper feed-forward block.

\section{Control schemes}

\subsection{Lambda controller tuning}

For the investigations presented here, lambda tuned PI controllers
will be used since both are commonly used in industry. The lambda
method \citep{panagopoulos1997lambda} is a two step procedure where
the first step is to approximate the transfer function by a first
order system with dead time, i.e,

\[
G^{*}(s)=\frac{K}{1+Ts}e^{-Ls}.
\]
Using the PI controller structure

\[
C(s)=K_{p}(1+\frac{1}{T_{i}s})
\]
the controller parameters are derived from $G^{*}(s)$ according to

\[
K_{p}=\frac{1}{K}\frac{T}{L+\lambda}
\]

\[
T_{i}=T
\]

\[
\lambda=\eta T,
\]
where $\lambda$ is the target time constant of the closed loop system,
and $\eta$ is a tuning parameter that will later be used to tune
$\lambda$.

\subsection{IMC controller}

An alternative to lambda tuned controllers, is to use IMC tuning,
which uses a model of the system to cancel out as much of the system
dynamics as possible. An IMC controller can be implemented in the
following way \citep{rivera1986internal}. Given a stable transfer
function model $\tilde{g}$ of the system, one starts by factorizing
the model into two parts:
\[
\tilde{g}=\tilde{g}_{+}\tilde{g}_{-}
\]
such that $\tilde{g}_{+}$ contains the delays and the non minimum
phase zeros of $\tilde{g}$, while $\tilde{g}_{-}$ contains the remaining
dynamics. This ensures that $\tilde{g}_{-}^{-1}$ is stable. A controller
can then be implemented as

\[
C=\frac{f\tilde{g}_{-}^{-1}}{1-f\tilde{g}_{+}}
\]
where,

\[
f=\frac{1}{(1+\epsilon s)^{q}}
\]
is a user designed filter, $\epsilon$ is a tuning parameter and $q$
is chosen such that the controller is proper. When implementing IMC
we will chose 

\[
\epsilon=\eta Z
\]
where $Z$ is the largest time constants of the model's non-minimum
phase zeros and $\eta$ is a tuning parameter. For minimum phase systems
we will instead chose $\epsilon$ as in lambda tuning, i.e.

\[
\epsilon=\eta T,
\]
where $T$ is the time constant of the system when approximated by
a first order system. 

\section{An illustrative example}

\begin{figure}
\includegraphics[scale=0.75]{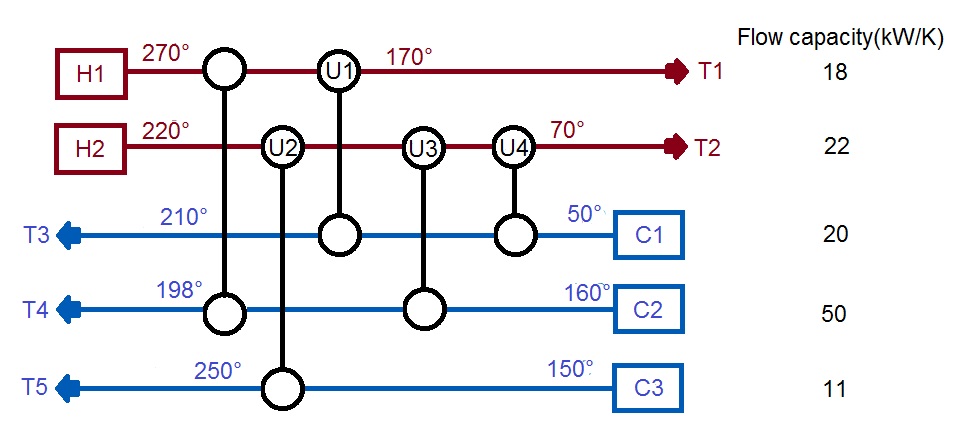}\caption{\label{fig:HEN-configuration-which}The studied heat exchanger network}
\end{figure}

To demonstrate some of the issues that can arise with the gramian
based measures, we examine a heat exchanger network (HEN), which is
a modified version of the configuration designed using pinch technology
in Case study 1 by \cite{Escobar2013801}. The modifications, implemented
to make the HEN interesting from a control configuration selection
perspective, amount to removing heaters and coolers and instead adding
one more heat exchanger (HE) connected to a new stream (C3). The modified
HEN can be seen in Figure \ref{fig:HEN-configuration-which}. The
specifications for streams H1-H2 and C1-C2 are the same as in  \cite{Escobar2013801},
and the new stream C3 has a flow capacity of 11 kW/K. The goal is
to control the output temperatures T1 to T4 using bypasses over the
heat exchangers U1 to U4. T5 is assumed to be controlled further downstream
and is thus not necessary to control here. 

\subsection{Heat exchanger models}

The heat exchangers are modeled as a series of mixing tanks (15 mixing
tanks were used to model U4, while 10 were found to be sufficient
for the other HEs), as described by the multi-cell model by \cite{mathisen1991controllability}.
Moreover, the heat transfer coefficients have been modified from those
in \cite{Escobar2013801} to compensate for not using the logarithmic
temperature differences, as discussed by \cite{mathisen1994dynamic}. 
All heat exchangers are modeled to have a residence time of 10 seconds.
Pipe residence time and heat losses are not included in the model. 

The actuators are controlled bypasses on the hot side of heat exchangers
U1 to U4. The HEs areas were therefore increased to retain the same
steady-state temperatures when they bypass 10 percent of the stream,
which consequently defines their stationary operating points. The
complete system model was implemented in Matlab/Simulink and linearised
with the bypass flow ratio on U1 to U4 as inputs and the temperatures
T1 to T4 as outputs. 

\subsection{The pairing problem}

On the linearised model, different input output pairing algorithms
were applied. The three previously mentioned gramian based methods were used to derive decentralized
control schemes. For comparison purposes we used the classical RGA \citep{bristol1966new} and the more
recent ILQIA \citep{Halvarsson309623} to derive alternative 
control schemes . The recommended pairing for each method is shown
in Table \ref{tab:The-results-for}, where we note that all the gramian
based methods suggest the same pairing different from the ILQIA and the RGA.
To compare the methods decentralized PI control schemes were implemented
and tuned using the lambda method applied on the open loop subsystems.
Each of the control configurations was simulated on the nonlinear system both with reference
steps on the output temperature of streams H1, H2, C1 and C2 and with
disturbances on the input temperature and flow rate of streams H1,
H2, C1 and C2. The size of the reference steps were two degrees and
the size of the disturbances on the input temperatures were negative
two degrees. For flow rate the system was tested with a decrease of
5\% on the flow rate of H1 and H2, and with an increase of the flow
rate of 5\% on C1 and C2. These disturbances were chosen to be of
a magnitude and direction that is possible for the system to completely
compensate for. The simulations ran for 1000 seconds after the reference
or disturbance step to fully observe its impact. For assessment, the
mean quadratic deviation from the reference was devised as a cost,
allowing comparison of the different pairing schemes. This was repeated
for controllers designed with different values of $\eta$, and the
result of those simulations are presented in Table \ref{tab:Costs}.

\begin{table}
\centering{}\caption{\label{tab:The-results-for}The results for the pairing suggestions
for the HEN}
\begin{tabular}{|c|c|c|c|c|c|}
\hline 
 & RGA & PM & HIIA & $\Sigma_{2}$ & ILQIA\tabularnewline
\hline 
T1 & U3 & U1 & U1 & U1 & U1\tabularnewline
\hline 
T2 & U4 & U4 & U4 & U4 & U4\tabularnewline
\hline 
T3 & U1 & U2 & U2 & U2 & U3\tabularnewline
\hline 
T4 & U2 & U3 & U3 & U3 & U2\tabularnewline
\hline 
\end{tabular}
\end{table}

{
\begin{table}
\centering{}\caption{\label{tab:Costs}Costs for different controller tuning $\eta$}
\footnotesize
\begin{tabular}{|c|c|c|c|}

\hline 
$\eta$ & RGA & Gramian based methods & ILQIA\tabularnewline
\hline 
1 & 235065 & 342149 & 263431\tabularnewline
\hline 
2 & 136606 & 210002 & 204696\tabularnewline
\hline 
3 & 95950 & 126502 & 156832\tabularnewline
\hline 
3,5 & 61002 & 83096 & 120940\tabularnewline
\hline 
4 & \textcolor{black}{28570} & 63220 & 101968\tabularnewline
\hline 
4,5 & 16528 & 41376 & 85213\tabularnewline
\hline 
5 & 7595 & 16443 & 69522\tabularnewline
\hline 
5,5 & \textbf{\textcolor{blue}{1788}} & \textcolor{black}{4771} & 53765\tabularnewline
\hline 
6 & 1966 & 3533 & 35830\tabularnewline
\hline 
6,5 & 2154 & 3282 & \textcolor{black}{21232}\tabularnewline
\hline 
7 & 2344 & \textbf{\textcolor{blue}{3235}} & \textcolor{black}{9054}\tabularnewline
\hline 
7,5 & 2537 & 3274 & 1570\tabularnewline
\hline 
8 & 2732 & 3358 & \textbf{\textcolor{blue}{990}}\tabularnewline
\hline 
10 & 3526 & 3897 & 1064\tabularnewline
\hline 
15 & 5568 & 5706 & 1687\tabularnewline
\hline 
\end{tabular}
\end{table}
}

A few conclusions can be drawn from these results. We can see that
for aggressive control schemes, all controller schemes fail, resulting
in an undamped oscillatory system. This is not unexpected as there
are obvious limits on the actuators (they cannot bypass more than
100\% of the stream or less than 0\%), and therefore the controllers
need to be somewhat cautious. However the control configuration suggested
by the gramian based methods yields a considerably worse control for
the best tuning than the ones recommended by the RGA or ILQIA, with
a minimum cost of 3235 as opposed to 1788 or 990 (table entries in
bold and blue). To examine why, we need to examine the IMs from the
gramian based methods:

{\footnotesize
\begin{eqnarray*}
PM & = & \begin{bmatrix}0.15 & 0.00046 & 0.056 & 0.13\\
0 & 0.000014 & 0.0023 & 0.55\\
0.058 & 0.00084 & 0.00052 & 0.017\\
0 & 0.0091 & 0.026 & 0
\end{bmatrix}\\
HIIA & = & \begin{bmatrix}0.16 & 0.0084 & 0.097 & 0.15\\
0 & 0.0015 & 0.018 & 0.29\\
0.1 & 0.011 & 0.009 & 0.054\\
0 & 0.032 & 0.063 & 0
\end{bmatrix}\\
\Sigma_{2} & = & \begin{bmatrix}0.17 & 0.000035 & 0.0054 & 0.0038\\
0 & 0.0000039 & 0.00081 & 0.81\\
0.0051 & 0.00003 & 0.0000072 & 0.00022\\
0 & 0.00036 & 0.00086 & 0
\end{bmatrix}.
\end{eqnarray*}
}
As can be seen, all the values in the second columns are small compared
to the largest values in the other columns. This means that when choosing
elements from the matrix little importance is given to the second
column. In other words, little importance is given to which element
should be controlled by U2. The reason for this is that U2 is not
particularly well suited for control compared to the other inputs,
and therefore the values in its column are lower. However, one can
still clearly see that the IMs suggest that U2 is much better suited
for controlling T4 than controlling T3, but as can be seen in Table
1, none of the gramian based measures recommend this pairing, while
both non-gramian based methods do. Similarly, we see that the third
and fourth rows contain considerably less interaction than the other
rows. Hence, less emphasis is placed on selecting a good actuator
for T3 and T4 compared to T1 and T2. 

It can be argued that this is a matter of scaling. However all the
inputs, being bypass percentages, are scaled from 0 to 1 as is the
general convention to resolve the issue of input scaling for the gramian
based methods \citep{salgado2004mimo}. Moreover, all the outputs
are tested with identically sized reference steps and can thus be
said to be properly scaled as well. However, in this case this scaling
scheme appears to be insufficient.  \cite{arranz2009new} suggest a scaling
for the $\Sigma_{2}$, where each element in the
IM is divided by the sum of all the elements in either its column
or row. This seems an attractive proposition to resolve this issue
as it ensures that either each input or each output is given equal
weight. If we scale the PM, $\Sigma_{2}$ and HIIA interaction measures
with this method, we get the configurations shown in Table \ref{tab:The-resultsscaled}.

\begin{table*}
\centering{}\caption{\label{tab:The-resultsscaled}The results for the optimal pairing
of HEN using the new scaling}
\begin{tabular}{|c|c|>{\centering}p{0.09\textwidth}|>{\centering}p{0.09\textwidth}|>{\centering}p{0.09\textwidth}|>{\centering}p{0.09\textwidth}|>{\centering}p{0.09\textwidth}|>{\centering}p{0.09\textwidth}|c|}
\hline 
 & {\scriptsize{}RGA} & {\scriptsize{}PM column scaling} & {\scriptsize{}HIIA column scaling} & {\scriptsize{}$\Sigma_{2}$ column scaling} & {\scriptsize{}PM row scaling} & {\scriptsize{}HIIA row scaling} & {\scriptsize{}$\Sigma_{2}$ row scaling} & {\scriptsize{}ILQIA}\tabularnewline
\hline 
{\scriptsize{}T1} & {\scriptsize{}U3} & {\scriptsize{}U3} & {\scriptsize{}U3} & {\scriptsize{}U1} & {\scriptsize{}U2} & {\scriptsize{}U2} & {\scriptsize{}U1} & {\scriptsize{}U1}\tabularnewline
\hline 
{\scriptsize{}T2} & {\scriptsize{}U4} & {\scriptsize{}U4} & {\scriptsize{}U4} & {\scriptsize{}U4} & {\scriptsize{}U4} & {\scriptsize{}U4} & {\scriptsize{}U4} & {\scriptsize{}U4}\tabularnewline
\hline 
{\scriptsize{}T3} & {\scriptsize{}U1} & {\scriptsize{}U1} & {\scriptsize{}U1} & {\scriptsize{}U3} & {\scriptsize{}U1} & {\scriptsize{}U1} & {\scriptsize{}U2} & {\scriptsize{}U3}\tabularnewline
\hline 
{\scriptsize{}T4} & {\scriptsize{}U2} & {\scriptsize{}U2} & {\scriptsize{}U2} & {\scriptsize{}U2} & {\scriptsize{}U3} & {\scriptsize{}U3} & {\scriptsize{}U3} & {\scriptsize{}U2}\tabularnewline
\hline 
\end{tabular}
\end{table*}
As can be seen in Table \ref{tab:The-resultsscaled}, with column
scaling we get the same control configurations as recommended by either
the RGA or the ILQIA and consequently a lowered cost according to
the assessment. With row scaling we get a new configuration for the
HIIA and PM\textbf{. }Testing with this configuration however yields
a minimum cost of 3227, so this configuration is not significantly
better than the unscaled gramian based configuration. 

To conclude, in this case row scaling yields no noteworthy improvement,
while column scaling yields a better configuration for all the tested
gramian based methods. 

\subsection{Scaling using the Sinkhorn-Knopp algorithm.}

While we could observe here that in this case, normalizing the IMs
columns such that each input was given equal weight worked well, while
normalizing the rows such that each output was given equal weight
yielded poorer results. There is another option, to ensure that
the IMs rows and columns all sum up to the same value, and hence all
inputs and outputs are given equal weight. This can be achieved by
iteratively alternating between scaling the elements by row sum and
by column sum, which is the Sinkhorn-Knopp algorithm described in
Section \ref{SKalgo}. Implementing the Sinkhorn-Knopp algorithm and stopping
the algorithm when the error is less than $10^{-3}$ yields the control
configurations presented in Table \ref{tab:The-resultsscaledSinkhorn}.

\begin{table}
\caption{\label{tab:The-resultsscaledSinkhorn}Optimal pairing of the HEN using
the Sinkhorn-Knopp algorithm}
\centering{}%
\begin{tabular}{|c|c|c|c|}
\hline 
 & PM  & HIIA  & $\Sigma_{2}$ \tabularnewline
\hline 
T1 & U3 & U3 & U3\tabularnewline
\hline 
T2 & U4 & U4 & U4\tabularnewline
\hline 
T3 & U1 & U1 & U1\tabularnewline
\hline 
T4 & U2 & U2 & U2\tabularnewline
\hline 
\end{tabular}
\end{table}
As can be seen, this resulted in the same configuration as the RGA,
which while not being the configuration with the lowest cost, still
yielded a considerably better result than the configuration recommended
by the unscaled gramian based measures. This is also the same configuration
obtained with the PM and HIIA when applying column scaling. However,
for the $\Sigma_{2}$ method column scaling yielded a configuration
with a lower cost. 

\section{Large scale assessment of the methods}

While the HEN example may have demonstrated potential benefits from
rescaling the IMs,a single case study is insufficient to draw general conclusions. To quantitatively assess the methods we have therefore
used the MIMO model generator described by \cite{Bengtsson2017}
to generate a large number of linear MIMO-systems, with specifications
shown in Table \ref{tab:MIMO-config}. The different scaling methods,
i.e. by row sum, column sum, or both using the SK algorithm, were
then applied to the generated models and the results were compared
to using only the standard scaling. In addition yet another approach
was implemented. In this approach one would either scale the rows
or the columns, according to the reasoning in Section 2.2.2.

\begin{table*}
\centering{}\caption{\label{tab:MIMO-config}Table showing the MIMO model generator\cite{Bengtsson2017} settings}
\begin{tabular}{|>{\raggedright}p{0.7\textwidth}|c|}
\hline 
Parameter & Default value\tabularnewline
\hline 
\textbf{Size} & \tabularnewline
\hline 
Number of inputs & 5\tabularnewline
\hline 
Number of outputs & 5\tabularnewline
\hline 
Minimum number of inputs affecting each output & 4\tabularnewline
\hline 
Maximum number of inputs affecting each output & 5\tabularnewline
\hline 
Minimum transfer function order & 1\tabularnewline
\hline 
Maximum transfer function order & 3\tabularnewline
\hline 
Minimum relative degree & 1\tabularnewline
\hline 
Maximum relative degree & 3\tabularnewline
\hline 
\textbf{Dynamics} & \tabularnewline
\hline 
Maximum static gain & 10-1000\tabularnewline
\hline 
Minimum pole time constant & 1\tabularnewline
\hline 
Maximum pole time constant & 10\tabularnewline
\hline 
Minimum damping for complex poles & 0.1\tabularnewline
\hline 
Distinct time constants & false\tabularnewline
\hline 
Basing zeros time constants on poles when possible & true\tabularnewline
\hline 
Maximum overshoot percentage & 10\tabularnewline
\hline 
Maximum undershoot percentage & 25\tabularnewline
\hline 
Tolerance when determining overshoot/undershoot & 0.01\tabularnewline
\hline 
Factor used to determine minimum time constant & 100\tabularnewline
\hline 
\textbf{Poles and Zeros} & \tabularnewline
\hline 
Maximum number of unstable poles & 0\tabularnewline
\hline 
Minimum number of unstable poles & 0\tabularnewline
\hline 
Maximum number of purely imaginary pole pairs & 0\tabularnewline
\hline 
Minimum number of purely imaginary pole pairs & 0\tabularnewline
\hline 
Percentage of unstable poles which are complex & 0\tabularnewline
\hline 
Percentage of stable poles which are complex & 20\tabularnewline
\hline 
Percentage of transfer functions with single integrators & 0\tabularnewline
\hline 
Percentage of transfer functions with double integrators & 0\tabularnewline
\hline 
Percentage of transfer functions with derivatives & 0\tabularnewline
\hline 
Maximum number of non-minimum phase zeros & 4\tabularnewline
\hline 
Minimum number of non-minimum phase zeros & 0\tabularnewline
\hline 
\textbf{Delay} & \tabularnewline
\hline 
Percentage of transfer functions with delay & 10\tabularnewline
\hline 
Minimum Delay & 0\tabularnewline
\hline 
Maximum Delay & 0.5\tabularnewline
\hline 
Pad{\'e} approximation order & 2\tabularnewline
\hline 
\end{tabular}
\end{table*}

For each control configuration a decentralized control scheme was
designed using both internal model control and the lambda method for varying values of $\eta$ (the
Matlab function \texttt{tfest} was used to approximate the transfer
functions as first order systems which were then used for the lambda controller
design). The entire feedback system was then tested both by reference
step and by load disturbance. For the comparison we define a cost
being the squared deviation from the reference for 2000 time units
after the reference step or the load disturbance. For each configuration
the cost was calculated for values of $\eta$ ranging from 0.1 to
10 and the lowest cost was then saved. Having calculated this cost
for each IM, each IM is given a score defined as

\[
S=\frac{c_{min}}{c}
\]
where $S$ is the score of the IM, $c$ is the IMs cost, and $c_{min}$
is the lowest cost of all IMs for the system. The score was set to
zero if the control scheme yielded unstable results. This measure
was chosen to normalize the score of each iteration between 0 and
1, to ensure that the results on different systems are comparable. 

Three sets of 150 randomly generated systems were assessed having
maximum static gains of 10, 100 and 1000 (minimum static gain was
always 1). Both decentralized and sparse control schemes (for $\rho=3$)
were implemented. The mean scores are presented in Table \ref{tab:tablescoreRef}
and Table \ref{tab:tablescoreD} for both $\lambda$-tuned controllers
and IMC controllers. 

The systems generated by the MIMO generator generally contained non-minimum phase transmission zeros. To evaluate how the presence of non-minimum phase dynamics affected the scaling methods, we also tested sets of system without non-minimum phase transmission zeros (still with specifications according to Table \ref{tab:MIMO-config}). The result of this is included in Tables \ref{tab:tablescoreRefMP} and \ref{tab:tablescoreDisMP}.

The large number of systems investigated allows a statistical evaluation
whether the new scalings yield statistically significant improvements
or not. Therefore, a t-test for paired samples was performed on the
hypothesis that the scaled systems had a higher score than the unscaled
system with a 95\% confidence interval. This evaluation was carried
out on both the systems with feed-forward and without. The statistically
significant improvements are highlighted in bold numbers in Tables
\ref{tab:tablescoreRef}-\ref{tab:tablescoreDisMP}.

\begin{table*}
\centering{}\caption{\label{tab:tablescoreRef}Average score for the different methods
with different gain variation for a reference step for non-minimum phase systems. Bold values indicate
statistically significant improvements compared to the system with
no scaling. }
\begin{tabular}{|l|c|c|c||c|c|c|}
\hline 
Maximum Gain & 10 & 100 & 1000 & 10 & 100 & 1000\tabularnewline
\hline 
\hline 
Controller design method & \multicolumn{3}{c||}{Lambda } & \multicolumn{3}{c|}{IMC}\tabularnewline
\hline 
PM &  &  &  &  &  & \tabularnewline
\hline 
\emph{Decentralized Control} &  &  &  &  &  & \tabularnewline
\hline 
No scaling & 0.45 & 0.38 & 0.30 & 0.38 & 0.45 & 0.35\tabularnewline
\hline 
Column scaling & \textbf{0.57} & \textbf{0.54} & \textbf{0.45} & \textbf{0.48} & \textbf{0.58} & \textbf{0.51}\tabularnewline
\hline 
Row scaling & \textbf{0.55} & \textbf{0.54} & \textbf{0.50} & \textbf{0.51} & \textbf{0.58} & \textbf{0.50}\tabularnewline
\hline 
Row/column scaling & \textbf{0.60} & \textbf{0.59} & \textbf{0.52} & \textbf{0.51} & \textbf{0.64} & \textbf{0.51}\tabularnewline
\hline 
Sinkhorn-Knopp scaling & \textbf{0.61} & \textbf{0.64} & \textbf{0.67} & \textbf{0.58} & \textbf{0.71} & \textbf{0.67}\tabularnewline
\hline 
\emph{Sparse Control} &  &  &  &  &  & \tabularnewline
\hline 
No scaling & 0.46 & 0.39 & 0.31 & 0.38 & 0.43 & 0.40\tabularnewline
\hline 
Column scaling & \textbf{0.59} & \textbf{0.61} & \textbf{0.58} & \textbf{0.46} & \textbf{0.57} & \textbf{0.58}\tabularnewline
\hline 
Row scaling & \textbf{0.56} & \textbf{0.52} & \textbf{0.47} & \textbf{0.47} & 0.50 & \textbf{0.48}\tabularnewline
\hline 
Row/column scaling & \textbf{0.62} & \textbf{0.62} & \textbf{0.60} & \textbf{0.50} & \textbf{0.63} & \textbf{0.58}\tabularnewline
\hline 
Sinkhorn-Knopp scaling & \textbf{0.61} & \textbf{0.64} & \textbf{0.67} & \textbf{0.58} & \textbf{0.71} & \textbf{0.67}\tabularnewline
\hline 
$\mathbf{\Sigma_{2}}$ &  &  &  &  &  & \tabularnewline
\hline 
\emph{Decentralized Control} &  &  &  &  &  & \tabularnewline
\hline 
No scaling & 0.42 & 0.37 & 0.24 & 0.41 & 0.40 & 0.33\tabularnewline
\hline 
Column scaling & \textbf{0.51} & \textbf{0.54} & \textbf{0.45} & \textbf{0.52} & \textbf{0.55} & \textbf{0.46}\tabularnewline
\hline 
Row scaling & \textbf{0.52} & \textbf{0.53} & \textbf{0.46} & \textbf{0.53} & \textbf{0.59} & \textbf{0.47}\tabularnewline
\hline 
Row/column scaling & \textbf{0.53} & \textbf{0.57} & \textbf{0.49} & \textbf{0.55} & \textbf{0.61} & \textbf{0.49}\tabularnewline
\hline 
Sinkhorn-Knopp scaling & \textbf{0.53} & \textbf{0.64} & \textbf{0.63} & \textbf{0.54} & \textbf{0.67} & \textbf{0.64}\tabularnewline
\hline 
\emph{Sparse Control} &  &  &  &  &  & \tabularnewline
\hline 
No scaling & 0.44 & 0.37 & 0.25 & 0.40 & 0.40 & 0.33\tabularnewline
\hline 
Column scaling & \textbf{0.53} & \textbf{0.61} & \textbf{0.52} & \textbf{0.51} & \textbf{0.57} & \textbf{0.51}\tabularnewline
\hline 
Row scaling & \textbf{0.50} & \textbf{0.51} & \textbf{0.47} & \textbf{0.52} & \textbf{0.53} & \textbf{0.41}\tabularnewline
\hline 
Row/column scaling & \textbf{0.54} & \textbf{0.61} & \textbf{0.52} & \textbf{0.56} & \textbf{0.60} & \textbf{0.49}\tabularnewline
\hline 
Sinkhorn-Knopp scaling & \textbf{0.53} & \textbf{0.64} & \textbf{0.63} & \textbf{0.54} & \textbf{0.67} & \textbf{0.65}\tabularnewline
\hline 
\textbf{HIIA} &  &  &  &  &  & \tabularnewline
\hline 
\emph{Decentralized Control} &  &  &  &  &  & \tabularnewline
\hline 
No scaling & 0.54 & 0.52 & 0.35 & 0.54 & 0.57 & 0.45\tabularnewline
\hline 
Column scaling & \textbf{0.60} & \textbf{0.61} & \textbf{0.55} & 0.58 & \textbf{0.66} & \textbf{0.58}\tabularnewline
\hline 
Row scaling & \textbf{0.62} & \textbf{0.60} & \textbf{0.58} & \textbf{0.59} & \textbf{0.67} & \textbf{0.56}\tabularnewline
\hline 
Row/column scaling & \textbf{0.65} & \textbf{0.63} & \textbf{0.60} & \textbf{0.59} & \textbf{0.69} & \textbf{0.59}\tabularnewline
\hline 
Sinkhorn-Knopp scaling & \textbf{0.64} & \textbf{0.66} & \textbf{0.67} & \textbf{0.62} & \textbf{0.72} & \textbf{0.67}\tabularnewline
\hline 
\emph{Sparse Control} &  &  &  &  &  & \tabularnewline
\hline 
No scaling & 0.53 & 0.54 & 0.37 & 0.54 & 0.56 & 0.45\tabularnewline
\hline 
Column scaling & \textbf{0.60} & \textbf{0.61} & \textbf{0.64} & 0.58 & \textbf{0.67} & \textbf{0.67}\tabularnewline
\hline 
Row scaling & \textbf{0.62} & \textbf{0.59} & \textbf{0.56} & \textbf{0.59} & 0.62 & \textbf{0.54}\tabularnewline
\hline 
Row/column scaling & \textbf{0.65} & \textbf{0.63} & \textbf{0.65} & \textbf{0.59} & \textbf{0.70} & \textbf{0.67}\tabularnewline
\hline 
Sinkhorn-Knopp scaling & \textbf{0.64} & \textbf{0.66} & \textbf{0.67} & \textbf{0.62} & \textbf{0.72} & \textbf{0.68}\tabularnewline
\hline 
\end{tabular}
\end{table*}

\begin{table*}
\centering{}\caption{\label{tab:tablescoreD}Average score for the different methods with
different gain variation for a load disturbance for non-minimum phase systems. Bold values indicate
statistically significant improvements compared to the system with
no scaling. }
\begin{tabular}{|l|c|c|c||c|c|c|}
\hline 
Maximum Gain & 10 & 100 & 1000 & 10 & 100 & 1000\tabularnewline
\hline 
\hline 
Controller design method & \multicolumn{3}{c||}{Lambda } & \multicolumn{3}{c|}{IMC}\tabularnewline
\hline 
\textbf{PM} &  &  &  &  &  & \tabularnewline
\hline 
\emph{Decentralized Control} &  &  &  &  &  & \tabularnewline
\hline 
No scaling & 0.46 & 0.42 & 0.36 & 0.35 & 0.51 & 0.46\tabularnewline
\hline 
Column scaling & \textbf{0.58} & \textbf{0.61} & \textbf{0.56} & \textbf{0.45} & \textbf{0.65} & \textbf{0.64}\tabularnewline
\hline 
Row scaling & \textbf{0.56} & \textbf{0.62} & \textbf{0.57} & \textbf{0.49} & \textbf{0.64} & \textbf{0.65}\tabularnewline
\hline 
Row/column scaling & \textbf{0.61} & \textbf{0.66} & \textbf{0.61} & \textbf{0.49} & \textbf{0.70} & \textbf{0.65}\tabularnewline
\hline 
Sinkhorn-Knopp scaling & \textbf{0.61} & \textbf{0.71} & \textbf{0.78} & \textbf{0.56} & \textbf{0.80} & \textbf{0.82}\tabularnewline
\hline 
\emph{Sparse Control} &  &  &  &  &  & \tabularnewline
\hline 
No scaling.  & 0.45 & 0.39 & 0.36 & 0.35 & 0.44 & 0.42\tabularnewline
\hline 
Column scaling & \textbf{0.56} & \textbf{0.59} & \textbf{0.57} & \textbf{0.43} & \textbf{0.57} & \textbf{0.57}\tabularnewline
\hline 
Row scaling & \textbf{0.54} & \textbf{0.53} & \textbf{0.53} & \textbf{0.45} & 0.51 & \textbf{0.51}\tabularnewline
\hline 
Row/column scaling & \textbf{0.60} & \textbf{0.63} & \textbf{0.62} & \textbf{0.47} & \textbf{0.61} & \textbf{0.56}\tabularnewline
\hline 
Sinkhorn-Knopp scaling & \textbf{0.61} & \textbf{0.71} & \textbf{0.78} & \textbf{0.56} & \textbf{0.80} & \textbf{0.82}\tabularnewline
\hline 
$\mathbf{\Sigma_{2}}$ &  &  &  &  &  & \tabularnewline
\hline 
\emph{Decentralized Control} &  &  &  &  &  & \tabularnewline
\hline 
No scaling.  & 0.45 & 0.42 & 0.29 & 0.44 & 0.48 & 0.45\tabularnewline
\hline 
Column scaling & \textbf{0.54} & \textbf{0.64} & \textbf{0.53} & \textbf{0.52} & \textbf{0.64} & \textbf{0.64}\tabularnewline
\hline 
Row scaling & \textbf{0.56} & \textbf{0.61} & \textbf{0.55} & \textbf{0.56} & \textbf{0.68} & \textbf{0.62}\tabularnewline
\hline 
Row/column scaling & \textbf{0.56} & \textbf{0.67} & \textbf{0.57} & \textbf{0.57} & \textbf{0.70} & \textbf{0.67}\tabularnewline
\hline 
Sinkhorn-Knopp scaling & \textbf{0.57} & \textbf{0.75} & \textbf{0.75} & \textbf{0.57} & \textbf{0.78} & \textbf{0.85}\tabularnewline
\hline 
\emph{Sparse Control} &  &  &  &  &  & \tabularnewline
\hline 
No scaling.  & 0.43 & 0.38 & 0.29 & 0.43 & 0.42 & 0.44\tabularnewline
\hline 
Column scaling & \textbf{0.55} & \textbf{0.63} & \textbf{0.54} & \textbf{0.50} & \textbf{0.57} & \textbf{0.56}\tabularnewline
\hline 
Row scaling & \textbf{0.51} & \textbf{0.52} & \textbf{0.51} & \textbf{0.54} & \textbf{0.52} & 0.48\tabularnewline
\hline 
Row/column scaling & \textbf{0.56} & \textbf{0.64} & \textbf{0.55} & \textbf{0.56} & \textbf{0.60} & \textbf{0.58}\tabularnewline
\hline 
Sinkhorn-Knopp scaling & \textbf{0.57} & \textbf{0.75} & \textbf{0.75} & \textbf{0.57} & \textbf{0.78} & \textbf{0.84}\tabularnewline
\hline 
\textbf{HIIA} &  &  &  &  &  & \tabularnewline
\hline 
\emph{Decentralized Control} &  &  &  &  &  & \tabularnewline
\hline 
No scaling.  & 0.55 & 0.57 & 0.43 & 0.52 & 0.62 & 0.58\tabularnewline
\hline 
Column scaling & \textbf{0.61} & \textbf{0.67} & \textbf{0.65} & 0.56 & \textbf{0.74} & \textbf{0.74}\tabularnewline
\hline 
Row scaling & \textbf{0.63} & \textbf{0.67} & \textbf{0.67} & \textbf{0.58} & \textbf{0.75} & \textbf{0.72}\tabularnewline
\hline 
Row/column scaling & \textbf{0.65} & \textbf{0.70} & \textbf{0.71} & \textbf{0.57} & \textbf{0.78} & \textbf{0.76}\tabularnewline
\hline 
Sinkhorn-Knopp scaling & \textbf{0.65} & \textbf{0.74} & \textbf{0.79} & \textbf{0.60} & \textbf{0.81} & \textbf{0.81}\tabularnewline
\hline 
\emph{Sparse Control} &  &  &  &  &  & \tabularnewline
\hline 
No scaling.  & 0.55 & 0.55 & 0.41 & 0.52 & 0.56 & 0.46\tabularnewline
\hline 
Column scaling & \textbf{0.61} & \textbf{0.66} & \textbf{0.66} & 0.56 & \textbf{0.69} & \textbf{0.64}\tabularnewline
\hline 
Row scaling & \textbf{0.63} & \textbf{0.64} & \textbf{0.60} & \textbf{0.58} & \textbf{0.66} & \textbf{0.60}\tabularnewline
\hline 
Row/column scaling & \textbf{0.65} & \textbf{0.70} & \textbf{0.70} & \textbf{0.57} & \textbf{0.75} & \textbf{0.68}\tabularnewline
\hline 
Sinkhorn-Knopp scaling & \textbf{0.65} & \textbf{0.74} & \textbf{0.79} & \textbf{0.60} & \textbf{0.81} & \textbf{0.81}\tabularnewline
\hline 
\end{tabular}
\end{table*}

\begin{table*}
\centering{}\caption{\label{tab:tablescoreRefMP} Average score for the different methods
with different gain variation for a reference step for minimum phase
systems. Bold values indicate statistically significant improvements
compared to the system with no scaling. }
\begin{tabular}{|l|c|c|c||c|c|c|}
\hline 
Maximum Gain & 10 & 100 & 1000 & 10 & 100 & 1000\tabularnewline
\hline 
\hline 
Controller design method & \multicolumn{3}{c||}{Lambda } & \multicolumn{3}{c|}{IMC}\tabularnewline
\hline 
\textbf{PM} &  &  &  &  &  & \tabularnewline
\hline 
\emph{Decentralized Control} &  &  &  &  &  & \tabularnewline
\hline 
No scaling & 0.24 & 0.27 & 0.22 & 0.26 & 0.28 & 0.26\tabularnewline
\hline 
Column scaling & \textbf{0.35} & \textbf{0.44} & \textbf{0.38} & \textbf{0.31} & \textbf{0.39} & \textbf{0.38}\tabularnewline
\hline 
Row scaling & \textbf{0.36} & \textbf{0.41} & \textbf{0.32} & \textbf{0.35} & \textbf{0.41} & \textbf{0.39}\tabularnewline
\hline 
Row/column scaling & \textbf{0.35} & \textbf{0.47} & \textbf{0.38} & \textbf{0.33} & \textbf{0.45} & \textbf{0.40}\tabularnewline
\hline 
Sinkhorn-Knopp scaling & \textbf{0.39} & \textbf{0.52} & \textbf{0.54} & \textbf{0.37} & \textbf{0.48} & \textbf{0.50}\tabularnewline
\hline 
\emph{Sparse Control} &  &  &  &  &  & \tabularnewline
\hline 
No scaling.  & 0.26 & 0.27 & 0.23 & 0.27 & 0.34 & 0.27\tabularnewline
\hline 
Column scaling & \textbf{0.35} & \textbf{0.51} & \textbf{0.49} & 0.31 & \textbf{0.50} & \textbf{0.47}\tabularnewline
\hline 
Row scaling & \textbf{0.36} & \textbf{0.44} & \textbf{0.39} & \textbf{0.35} & \textbf{0.44} & \textbf{0.41}\tabularnewline
\hline 
Row/column scaling & \textbf{0.34} & \textbf{0.52} & \textbf{0.47} & \textbf{0.34} & \textbf{0.54} & \textbf{0.46}\tabularnewline
\hline 
Sinkhorn-Knopp scaling & \textbf{0.39} & \textbf{0.52} & \textbf{0.55} & \textbf{0.37} & \textbf{0.48} & \textbf{0.50}\tabularnewline
\hline 
$\mathbf{\Sigma_{2}}$ &  &  &  &  &  & \tabularnewline
\hline 
\emph{Decentralized Control} &  &  &  &  &  & \tabularnewline
\hline 
No scaling.  & 0.66 & 0.46 & 0.34 & 0.64 & 0.50 & 0.40\tabularnewline
\hline 
Column scaling & 0.69 & \textbf{0.60} & \textbf{0.50} & \textbf{0.70} & \textbf{0.59} & \textbf{0.50}\tabularnewline
\hline 
Row scaling & \textbf{0.71} & \textbf{0.58} & 0.42 & \textbf{0.71} & \textbf{0.60} & \textbf{0.52}\tabularnewline
\hline 
Row/column scaling & \textbf{0.71} & \textbf{0.61} & \textbf{0.47} & \textbf{0.72} & \textbf{0.62} & \textbf{0.55}\tabularnewline
\hline 
Sinkhorn-Knopp scaling & 0.71 & \textbf{0.69} & \textbf{0.62} & \textbf{0.75} & \textbf{0.66} & \textbf{0.66}\tabularnewline
\hline 
\emph{Sparse Control} &  &  &  &  &  & \tabularnewline
\hline 
No scaling.  & 0.70 & 0.50 & 0.39 & 0.70 & 0.55 & 0.43\tabularnewline
\hline 
Column scaling & 0.74 & \textbf{0.73} & \textbf{0.63} & 0.73 & \textbf{0.76} & \textbf{0.69}\tabularnewline
\hline 
Row scaling & \textbf{0.77} & \textbf{0.66} & 0.51 & \textbf{0.77} & \textbf{0.66} & \textbf{0.54}\tabularnewline
\hline 
Row/column scaling & 0.74 & \textbf{0.73} & \textbf{0.58} & \textbf{0.76} & \textbf{0.76} & \textbf{0.63}\tabularnewline
\hline 
Sinkhorn-Knopp scaling & 0.71 & \textbf{0.69} & \textbf{0.62} & 0.75 & \textbf{0.66} & \textbf{0.66}\tabularnewline
\hline 
\textbf{HIIA} &  &  &  &  &  & \tabularnewline
\hline 
\emph{Decentralized Control} &  &  &  &  &  & \tabularnewline
\hline 
No scaling.  & 0.36 & 0.41 & 0.29 & 0.34 & 0.38 & 0.30\tabularnewline
\hline 
Column scaling & \textbf{0.41} & \textbf{0.51} & \textbf{0.43} & 0.36 & \textbf{0.46} & \textbf{0.44}\tabularnewline
\hline 
Row scaling & \textbf{0.46} & \textbf{0.47} & \textbf{0.39} & \textbf{0.39} & \textbf{0.44} & \textbf{0.40}\tabularnewline
\hline 
Row/column scaling & \textbf{0.42} & \textbf{0.51} & \textbf{0.45} & 0.37 & \textbf{0.47} & \textbf{0.47}\tabularnewline
\hline 
Sinkhorn-Knopp scaling & \textbf{0.45} & \textbf{0.55} & \textbf{0.51} & 0.38 & \textbf{0.50} & \textbf{0.51}\tabularnewline
\hline 
\emph{Sparse Control} &  &  &  &  &  & \tabularnewline
\hline 
No scaling.  & 0.36 & 0.43 & 0.44 & 0.34 & 0.41 & 0.35\tabularnewline
\hline 
Column scaling & \textbf{0.41} & \textbf{0.54} & \textbf{0.58} & 0.36 & \textbf{0.51} & \textbf{0.55}\tabularnewline
\hline 
Row scaling & \textbf{0.46} & 0.48 & 0.51 & \textbf{0.39} & 0.45 & \textbf{0.44}\tabularnewline
\hline 
Row/column scaling & \textbf{0.42} & \textbf{0.52} & \textbf{0.58} & 0.37 & \textbf{0.49} & \textbf{0.55}\tabularnewline
\hline 
Sinkhorn-Knopp scaling & \textbf{0.45} & \textbf{0.55} & 0.52 & 0.38 & \textbf{0.50} & \textbf{0.51}\tabularnewline
\hline 
\end{tabular}
\end{table*}

\begin{table*}
\centering{}\caption{\label{tab:tablescoreDisMP}Average score for the different methods
with different gain variation for a load disturbance for minimum phase
systems. Bold values indicate statistically significant improvements
compared to the system with no scaling.}
\begin{tabular}{|l|c|c|c||c|c|c|}
\hline 
Maximum Gain & 10 & 100 & 1000 & 10 & 100 & 1000\tabularnewline
\hline 
\hline 
Controller design method & \multicolumn{3}{c||}{Lambda } & \multicolumn{3}{c|}{IMC}\tabularnewline
\hline 
PM &  &  &  &  &  & \tabularnewline
\hline 
\emph{Decentralized Control} &  &  &  &  &  & \tabularnewline
\hline 
No scaling & 0.22 & 0.32 & 0.32 & 0.29 & 0.39 & 0.35\tabularnewline
\hline 
Column scaling & \textbf{0.35} & \textbf{0.52} & \textbf{0.59} & 0.32 & \textbf{0.53} & \textbf{0.53}\tabularnewline
\hline 
Row scaling & \textbf{0.34} & \textbf{0.47} & \textbf{0.52} & \textbf{0.37} & \textbf{0.55} & \textbf{0.54}\tabularnewline
\hline 
Row/column scaling & \textbf{0.35} & \textbf{0.55} & \textbf{0.60} & 0.34 & \textbf{0.60} & \textbf{0.57}\tabularnewline
\hline 
Sinkhorn-Knopp scaling & \textbf{0.38} & \textbf{0.58} & \textbf{0.78} & \textbf{0.38} & \textbf{0.63} & \textbf{0.73}\tabularnewline
\hline 
\emph{Sparse Control} &  &  &  &  &  & \tabularnewline
\hline 
No scaling & 0.23 & 0.30 & 0.33 & 0.29 & 0.37 & 0.34\tabularnewline
\hline 
Column scaling & \textbf{0.33} & \textbf{0.49} & \textbf{0.53} & 0.31 & \textbf{0.49} & \textbf{0.49}\tabularnewline
\hline 
Row scaling & \textbf{0.32} & \textbf{0.44} & \textbf{0.47} & \textbf{0.36} & \textbf{0.49} & \textbf{0.51}\tabularnewline
\hline 
Row/column scaling & \textbf{0.33} & \textbf{0.50} & \textbf{0.56} & 0.33 & \textbf{0.54} & \textbf{0.54}\tabularnewline
\hline 
Sinkhorn-Knopp scaling & \textbf{0.38} & \textbf{0.58} & \textbf{0.78} & \textbf{0.38} & \textbf{0.63} & \textbf{0.73}\tabularnewline
\hline 
$\mathbf{\Sigma_{2}}$ &  &  &  &  &  & \tabularnewline
\hline 
\emph{Decentralized Control} &  &  &  &  &  & \tabularnewline
\hline 
No scaling & 0.72 & 0.56 & 0.50 & 0.79 & 0.69 & 0.58\tabularnewline
\hline 
Column scaling & 0.77 & \textbf{0.76} & \textbf{0.75} & 0.82 & \textbf{0.82} & \textbf{0.75}\tabularnewline
\hline 
Row scaling & 0.77 & \textbf{0.72} & \textbf{0.68} & 0.83 & \textbf{0.78} & \textbf{0.75}\tabularnewline
\hline 
Row/column scaling & 0.78 & \textbf{0.76} & \textbf{0.73} & 0.84 & \textbf{0.82} & \textbf{0.81}\tabularnewline
\hline 
Sinkhorn-Knopp scaling & 0.78 & \textbf{0.86} & \textbf{0.89} & 0.83 & \textbf{0.86} & \textbf{0.92}\tabularnewline
\hline 
\emph{Sparse Control} &  &  &  &  &  & \tabularnewline
\hline 
No scaling & 0.69 & 0.55 & 0.48 & 0.65 & 0.63 & 0.56\tabularnewline
\hline 
Column scaling & \textbf{0.78} & \textbf{0.73} & \textbf{0.69} & \textbf{0.76} & \textbf{0.76} & \textbf{0.66}\tabularnewline
\hline 
Row scaling & 0.72 & \textbf{0.67} & \textbf{0.62} & 0.70 & 0.68 & \textbf{0.66}\tabularnewline
\hline 
Row/column scaling & \textbf{0.78} & \textbf{0.72} & \textbf{0.69} & \textbf{0.75} & \textbf{0.74} & \textbf{0.71}\tabularnewline
\hline 
Sinkhorn-Knopp scaling & \textbf{0.78} & \textbf{0.86} & \textbf{0.89} & \textbf{0.83} & \textbf{0.86} & \textbf{0.92}\tabularnewline
\hline 
\textbf{HIIA} &  &  &  &  &  & \tabularnewline
\hline 
\emph{Decentralized Control} &  &  &  &  &  & \tabularnewline
\hline 
No scaling & 0.34 & 0.48 & 0.45 & 0.35 & 0.49 & 0.42\tabularnewline
\hline 
Column scaling & \textbf{0.40} & \textbf{0.60} & \textbf{0.66} & 0.38 & \textbf{0.60} & \textbf{0.63}\tabularnewline
\hline 
Row scaling & \textbf{0.44} & \textbf{0.56} & \textbf{0.62} & \textbf{0.41} & \textbf{0.61} & \textbf{0.58}\tabularnewline
\hline 
Row/column scaling & \textbf{0.41} & \textbf{0.59} & \textbf{0.68} & 0.38 & \textbf{0.62} & \textbf{0.68}\tabularnewline
\hline 
Sinkhorn-Knopp scaling & \textbf{0.45} & \textbf{0.62} & \textbf{0.75} & 0.40 & \textbf{0.65} & \textbf{0.75}\tabularnewline
\hline 
\emph{Sparse Control} &  &  &  &  &  & \tabularnewline
\hline 
No scaling & 0.34 & 0.46 & 0.41 & 0.35 & 0.45 & 0.37\tabularnewline
\hline 
Column scaling & \textbf{0.40} & \textbf{0.59} & \textbf{0.60} & 0.38 & \textbf{0.58} & \textbf{0.60}\tabularnewline
\hline 
Row scaling & \textbf{0.44} & \textbf{0.55} & \textbf{0.59} & \textbf{0.41} & \textbf{0.56} & \textbf{0.52}\tabularnewline
\hline 
Row/column scaling & \textbf{0.41} & \textbf{0.59} & \textbf{0.66} & 0.38 & \textbf{0.60} & \textbf{0.64}\tabularnewline
\hline 
Sinkhorn-Knopp scaling & \textbf{0.45} & \textbf{0.62} & \textbf{0.74} & 0.40 & \textbf{0.65} & \textbf{0.75}\tabularnewline
\hline 
\end{tabular}
\end{table*}

As can be seen from the tables the scaled IMs fare considerably better
than the unscaled IMs. This improvement may be somewhat less pronounced
when the gain variation in the TFM is small which is in agreement
to what was observed in the HE example as high gain variation increases
the likelihood that a row or column in the IM will have considerably
less interaction than the other rows or columns. However, even in
the case of low gain variation there are statistically significant
improvements for many of the scaling schemes. 

One can also clearly see that the scaling method which yielded the
best results was the one based on the Sinkhorn-Knopp algorithm, as
it consistently yielded the highest score. The benefits of using the
Sinkhorn-Knopp algorithm were the most apparent with a high gain variation.
With a high gain variation one can see a very pronounced improvement
for the PM and $\Sigma_{2}$ method when using the Sinkhorn-Knopp
algorithm compared to the other scaling methods. When using the HIIA,
the improvement was somewhat smaller but still considerable. 

The scaled systems also fared better both for decentralized and sparse
control configurations. This indicates that the scaling methods can
be used for both. However the algorithm for implementing feedforward
was somewhat cautious, especially for Sinkhorn-Knopp scaling, as can
be seen in that there was often little difference between the cost
for the sparse and decentralized controllers. This also led to Sinkhorn-Knopp algorithm yielding comparably poor results in the cases where implementing feedforward had a very positive effect, namely reference following for minimum phase systems. A more thorough and
individual feedforward implementation for the different scaling methods might have alleviated this issue.

Generally the results are similar regardless whether the controllers
are tuned using IMC or the lambda method; in both cases the Sinkhorn-Knopp
algorithm generally outperformed the other scaling methods. 

Another observation is that without scaling it appears as if HIIA,
on average, gave better results than PM and $\Sigma_{2}$ for non-minimum phase systems, while the $\Sigma_{2}$ method was superior for minimuim phase systems. However, SK-scaling has an effect on the performance in many cases exceeding the difference between the different methods. 

Some caveats are however necessary. Only two methods for automatic
controller design was tested; it is possible that other control schemes
might yield different results. Furthermore, other than changing the
gain and the properties of the transmission zeros, only one set of model generator settings was used, modifications
to other system properties may possibly also yield different results.

\section{Conclusions and further work}

The gramian based interaction measures (IMs) are well known to be
affected by scaling and the standard is therefore to scale all inputs
and outputs to an equal range. A heat exchanger network control problem
illustrated a case where the conventional scaling scheme failed. It
was shown that by adapting the scaling of the IMs, this issue could
be resolved. A few possible scaling methods were tested on a large
number of systems using a MIMO model generator. It was found that
all the tested scaling methods statistically improved the performance
of the pairing methods consistently when designing a decentralized,
as well as a sparse controller. The scaling method that yielded the
best results was the one based on the Sinkhorn-Knopp algorithm, in
particular so when there are large variations in the static gains.
In addition, this scaling method has the advantage of yielding results
that are identical no matter what the original scaling of the inputs
and outputs is.

There is however some room for further research in this field. The automatic implementation of sparse controllers is something that can be explored further, especially when using Sinkhorn-Knopp scaling.

Morover there are other possible control strategies and controller tuning on which the scaling methods can be explored. When testing the scaling methods here we specifically examined the impact of non-minimum phase transmission zeros and differences in static gain. There are of course other system propeties which can be evaluated as well using the same techniques. 
\bibliographystyle{apacite}
\bibliography{example2.bib}

\end{document}